\documentclass[aps,pra,superscriptaddress,twocolumn,showpacs]{revtex4}
\usepackage{bm}
\usepackage{graphicx}
\usepackage{amsfonts}
\usepackage{amsmath}
\begin{document}

\title{Divergence of the entanglement range in low dimensional quantum systems}

\author{L. Amico}
\affiliation{MATIS-INFM $\&$ Dipartimento di Metodologie Fisiche e
Chimiche (DMFCI), Universit\`a di Catania, viale A. Doria 6, 95125
Catania, ITALY.}
\author{F. Baroni}
\affiliation{Dipartimento di Fisica dell'Universit\`a di Firenze,
via G.~Sansone 1, I-50019 Sesto F.no, ITALY.}
\author{A. Fubini}
\affiliation{MATIS-INFM $\&$ Dipartimento di Metodologie Fisiche e
Chimiche (DMFCI), Universit\`a di Catania, viale A. Doria 6, 95125
Catania, ITALY.}
\affiliation{Dipartimento di Fisica dell'Universit\`a di Firenze,
via G.~Sansone 1, I-50019 Sesto F.no, ITALY.}
\author{D. Patan\`e}
\affiliation{MATIS-INFM $\&$ Dipartimento di Metodologie Fisiche e
Chimiche (DMFCI), Universit\`a di Catania, viale A. Doria 6, 95125
Catania, ITALY.}
\author{V. Tognetti}
\affiliation{Dipartimento di Fisica dell'Universit\`a di Firenze,
via G.~Sansone 1, I-50019 Sesto F.no, ITALY.}
\affiliation{Istituto Nazionale di Fisica Nucleare, Sez. di Firenze,
via G.~Sansone 1, I-50019 Sesto F.no, ITALY.}
\author{Paola Verrucchi}
\affiliation{Centro di Ricerca e Sviluppo SMC dell'Istituto Nazionale di
Fisica della Materia-CNR, Sez. di Firenze, via G.~Sansone 1,
I-50019 Sesto F.no, ITALY.}

\date{\today}
\begin{abstract}
We study the pairwise entanglement close to separable ground states of
a class of one dimensional quantum spin models. At T=0 we find that such
ground states separate regions, in the space of the
Hamiltonian parameters, which are characterized by qualitatively different
types of entanglement, namely parallel and antiparallel entanglement; we
further demonstrate that the range of the Concurrence diverges while
approaching separable ground states, therefore evidencing
that such states, with uncorrelated fluctuations,
are reached by a long range reshuffling of the entanglement.
We generalize our results to the analysis of quantum phase transitions
occurring in bosonic and fermionic systems.
Finally, the effects of finite temperature are considered:
At $T{>}0$ we evidence the existence of a region where
no pairwise entanglement survives, so that entanglement, if
present, is genuinely multipartite.

\end{abstract}

% 75.10.Jm - Quantized spin models
% 05.30.-d - Quantum statistical mechanics
% 73.43.Nq - Quantum phase transitions
% 03.67.Mn - Entanglement production, characterization, and manipulation
\maketitle

\section{INTRODUCTION}

Quantum fluctuations may disorder the ground state of a system,
especially at low dimensions. Paradigmatic example in this sense are
quantum phase transitions~\cite{Sachdevbook99}, where different
phases can be achieved at $T=0$ by adjusting a control parameter of
the system. Paradoxically enough, quantum effects can provide also
classical-like ground states (CGS). In fact, for certain values of
the control-parameter, quantum fluctuations may become completely
uncorrelated; thereby the ground state of the system gets factorized
and identical to the lowest-energy state of the classical
counterpart of the original quantum system.  This phenomenon was
evidenced by Kurmann {\em et al.}~\cite{Kurmannetal82} in the early
eighties, for $S{=}1/2$ Heisenberg antiferromagnetic chains in an
external magnetic field $h$.

The study of entanglement in quantum many-body systems has been
providing a new angle in statistical
mechanics~\cite{Osterlohetal02,OsborneN02,Vidaletal03,VerstraeteMC04,CalabreseC04},
particularly at low temperature where cooperative phenomena are
dominated by quantum mechanics.  Thanks to a proper analysis of
certain entanglement estimators, the result by Kurmann {\it et al.}
was recently retrieved~\cite{Roscildeetal04} and generalized to
two-dimensional spin systems~\cite{Roscildeetal05}; moreover, numerical
evidences arose for it to hold in spin chains with long-ranged
interaction~\cite{DusuelV05}.

Special interest has been devoted to bipartite entanglement of formation
in connection with quantum criticality:
In fact, though quantum critical points are rather marked by the
enhancement of multipartite
entanglement~\cite{Roscildeetal04,Roscildeetal05,Anfossietal05},
the variation of the pairwise entanglement at criticality captures the
non analyticity  of the ground state of the system~\cite{WuSLS05}.
On the other hand, the naive guess that the range of pairwise entanglement
should diverge at a quantum phase transition, in analogy with the
divergence of the correlation length of the two-point correlators,
has never been evidenced~\cite{Osterlohetal02,OsborneN02}.

In this paper we show that, in the space of the Hamiltonian parameters,
the special points where CGS occur
(hereafter called {\it separable} points) mark a separation between
regions characterized by different types of entanglement,
(called  antiparallel and parallel entanglement in
Ref.~\cite{Fubinietal06}),
which correspond to qualitatively different spin configurations.
The transition from one region to the other is found to be characterized
by the divergence of the range of pairwise entanglement in the immediate
neighborhoods of the CGS (see Eqs.~(\ref{e.divRXY}) and (\ref{e.divXXZ})
below).

Spin-off in systems of strongly interacting bosons are also found: We
evidence that the superfluid-insulator quantum phase transition at commensurate
filling is in fact a transition between a phase (superfluid) where solely
particle-hole entanglement is present and a phase (insulator) with no pairwise
entanglement at all: the particle-hole entanglement is easily seen to correspond
to antiparallel entanglement and the range of the concurrence is found to
diverge while approaching the transition (see the paragraph below
Eq.~(\ref{e.divXXZ})).

Finally, we study how robust CGS are with respect to temperature: We
evidence the emergence of a region in the $h-T$ plane, fanning out from
separable points (see Fig.~\ref{f.fig2}) where pairwise entanglement
vanishes. In such region, if entanglement is present in the system, it
necessarily is of multipartite type.
Our study does also shed light on the result by Kurmann {\it et al.}
(see the concluding paragraph).

\section{MODELS}

We focus our attention on the class of one-dimensional spin models described
by the Hamiltonian
\begin{align}
{\cal{H}}(j_x,j_y,j_z)=J\sum_i&\big(j_x S_i^x S_{i+1}^x {+} j_y
S_i^y
S_{i+1}^y {+} j_z  S_i^z S_{i+1}^z \nonumber\\
& -h S_i^z\big)~,
\label{e.H}
\end{align}
where $i$ runs over the sites of the chain, $S_i^\alpha$ ($\alpha=x,y,z$)
are quantum angular momentum operators corresponding to $S=1/2$,
$j_\alpha$ are the anisotropy parameters ($|j_\alpha|\leq1$),
$h\equiv g\mu_{\rm B}H/J$ is the reduced magnetic field, and
$J$ is the exchange integral, assumed positive and hereafter set to unity.

In Ref.\onlinecite{Kurmannetal82} it was demonstrated that CGS are
obtained for $h=h_{\rm f}\equiv\sqrt{(j_x+j_z)(j_y+j_z)}$;
although the model (\ref{e.H}) cannot be solved exactly for generic
$j_\alpha$, the above result is rigorous. In order to analyze quantum
correlations, which are crucial for understanding the behavior of the
system at and near a separable point, we restrict the parameters in the
Hamiltonian (\ref{e.H}) so as to rely on exact results: We therefore
consider the solvable cases ${{\cal{H}}(1+\gamma,1-\gamma,0)}
\doteq {\cal H}_{XY}$ with $0\leq\gamma\leq 1$,  and
${{\cal{H}}(1,1,j_z)}\doteq {\cal H}_{XXZ}$ corresponding to the $XY$
and $XXZ$ models in a transverse field, respectively.

The quantitative analysis of the pairwise entanglement between two
spins sitting on sites $l$ and $m$ of the chain is
addressed via the concurrence $C_r$ with $r\equiv|l-m|$ (translational
invariance is assumed)~\cite{Wootters98}. In terms of
correlation functions
$g_r^{\alpha\alpha}=\langle S^\alpha_l S^\alpha_{l+r} \rangle$
and magnetization $M_z=\langle S^z \rangle$,
the concurrence reads~\cite{Amicoetal04}
\begin{eqnarray}
C_r&=&2\max\{0,C'_r,C''_r\}~,
\label{e.Cr}\\
C'_r&=&|g_r^{xx}+g_r^{yy}|-\sqrt{\left(\frac{1}{4}+g^{zz}_r\right)^2-M_z^2}~,
\label{e.C'r}\\
C''_r&=&|g_r^{xx}-g_r^{yy}|+g_r^{zz}-\frac{1}{4}~,
\label{e.C''r}
\end{eqnarray}
where $C'_r$ and $C''_r$ measure the pairwise entanglement related
with the occurrence of antiparallel and parallel Bell states,
respectively, as discussed in Ref.\onlinecite{Fubinietal06} both for
pure and mixed states.  We will also consider the
one-tangle~\cite{MeyerW02,Amicoetal04}
\begin{equation}
\tau_1=1-4\sum_\alpha M_\alpha^2~,
\end{equation}
the two-tangle
\begin{equation}
\tau_2=2\sum_r C^2_r~,
\end{equation}
and the ratio
$\tau_2/\tau_1$ which estimates the fraction of the total entanglement
stored in pairwise correlations~\cite{Roscildeetal04}.  Although
$g_r^{\alpha\alpha}$ and $M_z$ do not show any anomaly at a separable
point, they unveil it when combined in $C_r$, which is found to drop
to zero in a non-analytic way at this point. Such circumstance comes
together with the vanishing of $\tau_1$, and with a highly non-trivial
behavior of the ratio $\tau_2/\tau_1$~\cite{Roscildeetal04}.

\section{RESULTS}

\subsection{Zero Temperature}

Let us first consider the completely
integrable~\cite{LiebMS61,Niemeijer67} $XY$
model in a transverse field. Besides the quantum critical point $h=h_c=1$,
its $T=0$ phase diagram~\cite{BarouchMD71} is characterized
by the circle $h^2+\gamma^2=1$ (hereafter called the {\it separable
circle}) where CGS occur in the form
\begin{eqnarray*}
|GS^{XY}>&=&\prod_i|\phi^{XY}_i>~,\\
|\phi^{XY}_i>&\equiv&
(-1)^i\cos(\frac{\theta_\gamma}{2})|\downarrow_i>
     +\sin(\frac{\theta_\gamma}{2})|\uparrow_i>~,\\
\cos(\theta_\gamma)&=&\sqrt{\frac{1-\gamma}{1+\gamma}}~,
\end{eqnarray*}
where $|\phi^{XY}_i>$ is the state of the spin sitting on the $i$-th site.

Since the early papers on the model it is known that the functional
form of $M_z$ and $g^{\alpha\alpha}_r$ depends substantially on
whether the parameters of the system fix a point sitting inside,
outside, or on the circle itself~\cite{BarouchMD71}; however, it had never
been noticed that the simplicity of the exact solution at
$h_{\rm f}^{XY}=\sqrt{1-\gamma^2}$ is ultimately due
to the factorization of the ground state.
In fact, this is clearly evidenced by the concurrence, whose expression on
the separable circle reads
\begin{equation}
C_r=2C'_r=2C''_r=\frac{\gamma}{1+\gamma}+2M_z^2-\frac{1}{2}~~~~\forall
r~,
\label{e.C'onthecircle}
\end{equation}
and hence,
being $M_z=\frac{1}{2}\sqrt{(1-\gamma)/(1+\gamma)}~$,
\begin{equation}
C_r=C'_r=C''_r=0~,~~~\forall r~.
\label{e.nullC'onthecircle}
\end{equation}
Moreover, it is $C'_r\geq 0$ (and $C''_r\leq 0$) inside the circle,
and $C'_r\leq 0$ (and $C''_r\geq 0$) outside the circle,
no matter the sign of the exchange interaction and the
value of $r$. According to the analysis presented in
Ref.\onlinecite{Fubinietal06} this means that inside (outside) the circle
pairwise
entanglement exclusively originates from the occurrence of antiparallel
(parallel) Bell states.

\begin{figure}
\includegraphics[width=75mm]{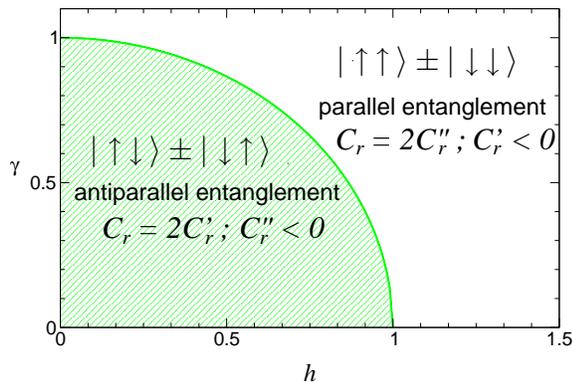}
\caption{\label{f.fig1a}
Entanglement phase diagram in the $h-\gamma$ plane. Points on the circle
correspond to models whose ground state is separable.}
\end{figure}

We remark that whether the system has parallel or antiparallel pairwise
entanglement at $T=0$ uniquely
depends on the Hamiltonian parameters $\gamma$ and $h$: As a consequence,
we can draw an ``entanglement phase diagram'' in the $h-\gamma$ plane,
where different phases are characterized by the presence of parallel or
antiparallel entanglement.  The separable circle $h^2+\gamma^2=1$
marks a boundary in such diagram (see Fig.~\ref{f.fig1a}) suggesting that
the occurrence of a CGS is a necessary step for switching from parallel to
antiparallel entanglement.
We notice that the same scenario emerges from the numerical analysis of
more complicated models, both in one~\cite{Roscildeetal04} and two
dimensions~\cite{Roscildeetal05}.

As the transition from parallel to antiparallel entanglement involves the
whole system, we study how the pairwise entanglement propagates along the
chain in the vicinity of a separable point: for doing that, we introduce
the range $R$ of the $T=0$ pairwise entanglement, which is
defined as the maximum distance between two spins such that the concurrence
is non vanishing:
\begin{equation}
C_r>0 ~~~{\rm for}~~r\leq R ~~~{\rm and}~~ C_r=0 ~~~{\rm for}~~ r>R~.
\label{e.Rdef}
\end{equation}

\begin{figure}
\includegraphics[width=75mm]{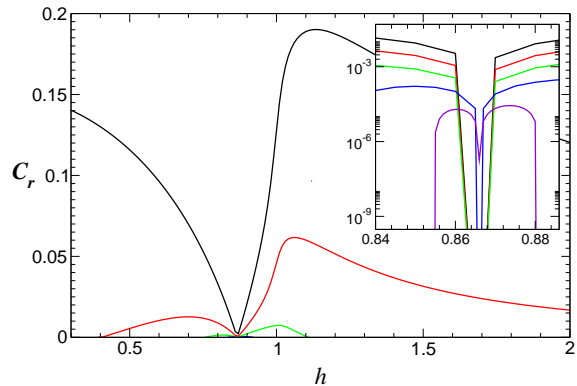}
\caption{\label{f.fig1b}
Concurrence for the XY model at $T=0$: curves are
$C_r$ versus $h$ for $r=1,2,3,4,5$ (top to bottom), and
$\gamma=0.5$. The inset is a zoom near the factorizing field, in
logarithmic scale.}
\end{figure}

\noindent We underline that the exact vanishing of $C_r$ for $r>R$ and
$h\neq h_{\rm f}$  follows from the definition of the concurrence
Eq.~(\ref{e.Cr}), in the sense that $C_r$ vanishes whenever $C'_r$ and
$C''_r$ are both negative. On the other hand, at the factorizing field,
$C_r=C_r'=C_r''=0$ for all values of $r$ due to the fact that the correlation
functions do not depend upon $r$ on the separable circle. In
Fig.~\ref{f.fig1b} we show the exact results for $C_r$ with $r$
ranging from 1 to 5. Results for larger $r$ are also available and show
the same qualitative behaviour: $C_r$ fans out from the separable point
with non-zero derivative, reaches a maximum, and then vanishes, both for
$h>h_f$ and $h<h_f$, though not simmetrically with respect to $h_f$.

The observed behaviour suggests the divergence of $R$ for $h\to h_{\rm
f}$: by expanding the exact expressions of the correlation functions up
the first order in $(h-h_{\rm f})$ we find~\cite{BaroniTesi06}
\begin{equation}
C_r=\frac{\Gamma^{2r-1}}{2\gamma}|h-h_{\rm f}|+{\cal{O}}((h-h_{\rm
f})^2)~, \label{e.Crexp1}
\end{equation}
where $\Gamma\equiv\sqrt{(1-\gamma)/(1+\gamma)}$. Eq.~(\ref{e.Crexp1})
confirms that all $C_r$ get progressively positive for $h$ approaching the
factorizing field, as clearly seen in Fig.~\ref{f.fig1b}: This means that
the range of the concurrence $R$ diverges at $h_{\rm f}$.

In order to analyze how $R$ diverges with the field, we push forward the
expansion in $(h-h_{\rm f})$, meanwhile considering the
large-$R$ expressions for the correlation
functions, given in Refs.\onlinecite{BarouchMD71} and \onlinecite{ItsJK05}.
The result for $h>h_{\rm f}$ reads
\begin{equation}
C''_r=\frac{\Gamma^{2r-1}}{4\gamma}(h-h_{\rm f})-[A^2+B(r)](h-h_{\rm
f})^2+{\cal{O}}((h-h_{\rm f})^3)~ \label{e.Crexp2}
\end{equation}
where $A^2\equiv \Gamma^2(3+\gamma)/(32\gamma^3)$ and $B(r)$ is a
coefficient which vanishes for $r\rightarrow\infty$. The range of
the concurrence is implicitely obtained from Eq.~(\ref{e.Crexp2}) by
requiring $C''_R=0$. In fact, for sufficiently large $r$, it is
$(A^2+B(r))^{-1}\simeq A^{-2}(1-B(r)/A^2)$, and hence, by
substituting into Eq.~(\ref{e.Crexp2}), $C''_r$ is found to vanish
both for $h=h_{\rm f}$ and $h-h_{\rm f}=\Gamma^{2r-1}/(4\gamma
A^2)$, leading to the logarithmic divergence
\begin{equation}
R^{XY}\propto \left (\ln{\frac{1-\gamma}{1+\gamma}} \right )^{-1}
\ln{\left| h-h_{\rm f} \right |^{-1}}~,
\label{e.divRXY}
\end{equation}
where we have introduced the symbol $R^{XY}$ to make clear that the
functional form of the divergence is in general model-dependent.

For $h<h_{\rm f}$ the expression for $C'_r$ is different from
Eq.~(\ref{e.Crexp2}), and to this difference we ascribe the
asymmetric behaviour of $C_r$ with respect to the separable point
observed in Fig.~\ref{f.fig1b}; however, for
$h\rightarrow h_{\rm f}^-$ the above result is still valid, though just for
the (thermal) ground state with unbroken symmetry~\cite{OsborneN02,Syljuasen03}.
It is to be noticed that, in the thermodynamic limit (here considered)
and while approaching the CGS, the fact that
the concurrences $C_r$ become progressively positive for larger
and larger $r$ goes together with their getting vanishingly
small: this is due both to the monogamy of the
entanglement~\cite{CoffmanKW00,OsborneV05} and to the
proximity of the separable point itself.

The divergence of $R$ implies, as a consequence of the monogamy of
the entanglement, that the role of pairwise
entanglement is enhanced while approaching the separable point; in
fact, the ratio $\tau_2/\tau_1$ is found to have a cusp minimum at
the critical point and to increase while moving
towards the CGS, in full analogy with the result of
Refs.\onlinecite{Roscildeetal04} and \onlinecite{Roscildeetal05}.
In the particular case of the Ising model
(i.e. $\gamma=1$), we find that for $h \rightarrow h_{\rm f}=0$
the ratio $\tau_2/\tau_1$ goes to unity at the
separable point.
For $\gamma\neq 1$ and $h_{\rm f}<h<h_c$ our data show that
$\tau_2/\tau_1$ monotonically increases for $h\rightarrow h_{\rm f}^+$
and that the value $(\tau_2/\tau_1)|_{h_{\rm f}^+}$
increases with $\gamma$.

We remark that the divergence of $R^{XY}$ while approaching the
separable circle cannot be recognized as a critical effect, since the
ground state is non-singular at $h_{\rm f}$ and the long-ranged
pairwise entanglement does not survive neither inside nor outside the
separable circle.

A complementary view on the physics of CGS is
obtained by the analysis of ${\cal H}_{XXZ}$, that can be done
resorting to Bethe ansatz results~\cite{Takahashibook99}. In this case
$h_{\rm f}$ coincides with the saturation field
$h_{\rm s}=(1+j_z)$, and $|GS^{XXZ}>=\prod_i|\uparrow_i>$.
Also, and distinctively from the $XY$ case, the factorized state extends
over a finite portion of the $h-j_z$ phase diagram of the model.
In fact, $h_{\rm s}$ separates a gapless {\it quasi-ordered} phase (with
power law decay of the in-plane correlation functions) from the gapped, fully
polarized phase (with $\langle S_z\rangle=1/2$).
Due to the in-plane symmetry of the model
(implying $g^{xx}_r=g^{yy}_r$) $C''_r$ is always negative and therefore
pairwise entanglement, if present, is of antiparallel type.
Combining the exact results of Refs.\onlinecite{Jinetal04}
and~\onlinecite{Hikiharaetal04}, we find that $R$ diverges
approaching the band transition as
\begin{equation}
R^{XXZ}\propto (h_{\rm s}-h)^{-{\theta}/{4}}~,
\label{e.divXXZ}
\end{equation}
where
\begin{equation}
\theta=2+\frac{4\sqrt{h_{\rm s}-h}}{\pi\tan(\frac{\pi\eta}{2})\tan(\pi\eta)}
~,~{\rm and}~~~\eta=\frac{1}{\pi}{\rm arcos}(-j_z).
\end{equation}
The divergence of $R$ in the isotropic case ($j_z=1$),
specifically determined in Ref.\onlinecite{Fubinietal06},
results from Eq.~(\ref{e.divXXZ}) with $\theta=2$.
We underline that saturation is not related to a spontaneous symmetry
breaking, and the divergence of $R$ while approaching $h_{\rm s}$ does not
mark a critical phenomenon.

Let us now consider strongly interacting
hard-core-bosons/spinless-fermions in
$1d$: We shall find that the occurence of CGS plays a fundamental role for
such systems.

Hard-core bosons with repulsive Coulomb interaction are described by
the quantum lattice gas~\cite{YangY66}, whose dynamics is described by
${\cal H}_{XXZ}$ phrasing the spins in terms of hard-core bosonic
operators: $a=S^-$, $a^\dagger=S^+$, $a^\dagger a=S^z+1/2$. The
relevant energies $t{\rightarrow}j_x$ (here $j_x{=}j_y{=}1$),
$U{\rightarrow}j_z$, and $\mu{\rightarrow}h+j_z$ are the hopping
amplitude, the Coulomb interaction, and the chemical potential,
respectively.  By this mapping, the superfluid and insulating
behaviors of the quantum lattice gas at commensurate filling are related to the quasi-ordered (partially
filled band)
and fully-polarized (filled band) phase of the XXZ spin model, respectively;
therefore, the {\em band transition} observed at $\mu{=}t{+}2U$, corresponds
to saturation occurring at $h_{\rm s}{=}(1{+}j_z)$.

Based on the
above analysis, we state that the insulator-superfluid band
transition is characterized by the divergence of the range of the
concurrence. Remarkably, the antiparallel character of the pairwise
entanglement present in the $XXZ$ model reflects the fact that close
to the superfluid-insulator transition exclusively particle-hole
entanglement plays a role.  Arguments along the same line can be
applied to spinless fermions models obtained via Jordan-Wigner
transformation of ${\cal H}_{XXZ}$
\cite{MikeskaK04}: The band-transition there observed is from an
insulating regime to a gapless phase equivalent to a Luttinger
liquid.

\subsection{Finite Temperature}

\begin{figure}
\includegraphics[width=75mm]{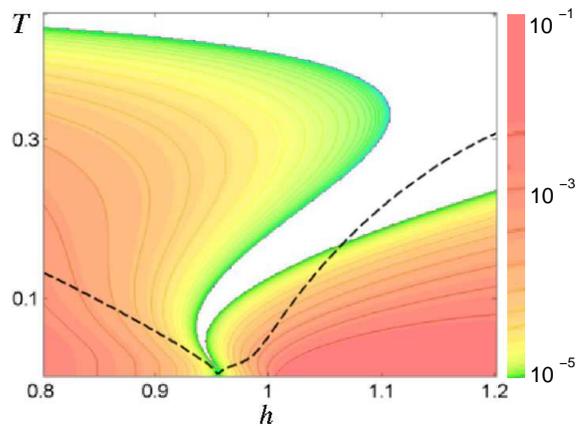}
\caption{Contour plot of $\tau_2$ in the $h-T$ plane, for $\gamma=0.3$
(i.e. $h_f=0.9539...$). The white area indicates where
$\tau_2=0$. Condition (\ref{e.witness}) is fullfilled below the
dashed line, which is defined by $\langle{\cal H}\rangle=E_{\rm sep}$.}
\label{f.fig2}
\end{figure}

\begin{figure}
  \includegraphics[width=80mm]{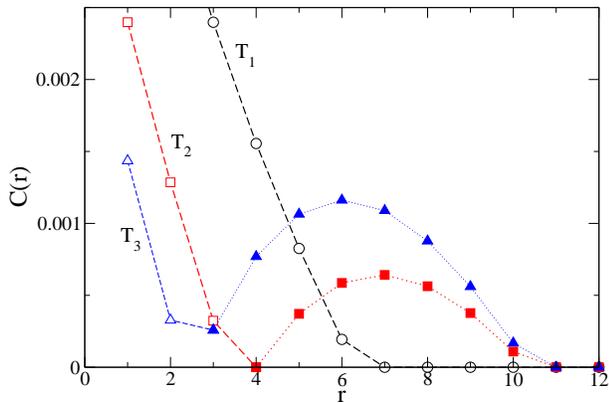}
\caption{$C_r$ versus $r$ for $\gamma=0.01$ and $h=1.1$.
Circles, squares and triangles correspond to three different temperatures:
$T_1{=}4\,10^{-3}$, $T_2{=}4.7\,10^{-3}$, and $T_3{=}5\,10^{-3}$,
respectively.
Full symbols mean $C_r{=}C'_r$ and empty symbols mean $C_r{=}C''_r$.}
\label{f.fig3}
\end{figure}

We now switch on a finite temperature in the system.  We consider
questions like: What is the effect fanning out from CGS on the thermal
(mixed) states of the system? Particularly: How meaningful is the
characterization of the system in terms of parallel or antiparallel
pairwise entanglement for mixed states? To answer these questions we
consider the $XY$ model where both parallel and antiparallel
entanglement are present at $T=0$. The analysis of $\tau_2$ evidences that in
the $h-T$ plane
it exists a region, fanning out from the CGS, where pairwise
entanglement vanishes (white region in Fig.~\ref{f.fig2}) and the
entanglement, if present, is shared between three or more parties.
This
means that a CGS may evolve into a quantum mixed state with genuinely
multipartite entanglement {\em by increasing temperature}.
In order to determine whether this happens or not, we need to know if
entanglement is present in the system: the one-tangle, which accomplishes
this task at $T=0$, is not a proper estimator
for the entanglement content of the system at finite temperature;
therefore, we have
to refer to some entanglement witness.
Following the results by T\'oth~\cite{Toth05}, entanglement is present if
\begin{equation}
\label{e.witness}
\langle {\cal H} \rangle - E_{\rm sep} < 0~,
\end{equation}
where $E_{\rm sep}$ is the
ground state energy of the corrisponding
classical model.
The region below the dashed line in Fig.~\ref{f.fig2} is where condition
\ref{e.witness} is fulfilled, i.e. where entanglement of whatever type is
present in the system.
We further observe that, in contrast to the analysis of the ground state, at
finite temperature we cannot characterize the two separate phases of
parallel and antiparallel entanglement. In fact, the two types of
entanglement (though well defined also for mixed states) can swap by varying
$T$ and/or $r$ (see Fig.~\ref{f.fig3}). The exchange between parallel
and antiparallel entanglement occurs in a non trivial way, that
ultimately produces the temperature ``reentrance'' of $\tau_2$ seen in
Fig.\ref{f.fig2}. We also find a regime where $C_r$ can be a non
monotonic function of $r$, so that, for instance, one spin is not
entangled with its nearest neighbor while being entangled with its
next-nearest neighbor. Such situations occur in the vicinity of the
region where $C_r$ vanishes, as seen in Fig.~\ref{f.fig3} and it is
due to the non-monotonic behavior\cite{BarouchMD71} of the correlation
functions.

\section{CONCLUSIONS AND PERSPECTIVES}

Summarizing, we have studied the occurrence of CGS in relation with
pairwise entanglement, in a class of $S=1/2$ spin chains. Our
results show that at $T=0$ the space of the Hamiltonian parameters is
divided into regions where either exclusively parallel or exclusively
antiparallel pairwise entanglement is present, no matter the distance
between the considered spins: Therefore, an entanglement phase
diagram can be unambiguously drawn. Transition lines in such diagram
corresponds to separable ground states, and are further
characterized by the fact that the range of the concurrence diverges while
moving towards them. Due to the monogamy of the entanglement, such divergence
goes together with the asymptotic vanishing of the value of the concurrence
itself.

We further provide (to our knowledge for the first time) an
explanation of the phenomenon described by Kurmann {\it et
al.}\cite{Kurmannetal82}: the factorization of the ground state is a
necessary step for antiparallel entanglement to be fully replaced by
parallel entanglement.
% i.e. for entanglement phase transition to occur.
We observe that in the global reshuffling of the ground
state which leads to a CGS, and involves all the spins of the chain, a
long range entanglement appears as a crucial ingredient.
Moreover, our results for the entanglement ratio $\tau_2/\tau_1$
confirm that multipartite entanglement dominates at a genuine quantum critical
point, while pairwise entanglement is essential for understanding the
mechanism leading to CGS in quantum systems.

Our analysis is of relevance also for bosonic systems undergoing
a superfluid-insulator transition. It is intriguing to conjecture
that the divergence of the range of $C_r$ at a CGS goes beyond model
dependency. In fact, Anfossi et al.~\cite{Anfossietal05} have observed a similar
divergence of the range of  $C_r$ also in the bond-charge extended Hubbard
model at certain transition lines.

For finite temperature the entanglement in the system
cannot be characterized by the single parameter  $h$ and a much
more complicated scenario emerges:
In particular we find that, by increasing $T$, the factorized (pure)
ground state evolves into a thermal (mixed) state with null pairwise
entanglement: this opens the possibility for the existence of an
experimentally accessible finite-temperature region where entanglement, if
present, is genuinely multipartite.

Finally, we notice that the possibility of controlling
via a proper tuning of the external magnetic field whether two
spins are entangled or not, and whether they share parallel or antiparallel
entanglement, might be of interest both from the experimental point of
view and for applicative purposes.

\section{Aknowledgements}

P.V. wishes to thank S~.Bose and D.~Burgarth for valuable discussions.
Comments by T.~Roscilde are gratefully aknowledged. This work sets in the
framework of the PRIN2005029421 project.

\end{document}